# How Carbon Border Adjustment Mechanism is Energizing the EU Carbon Market and Industrial Transformation

By Joseph Nyangon, PhD and Brecht Seifi

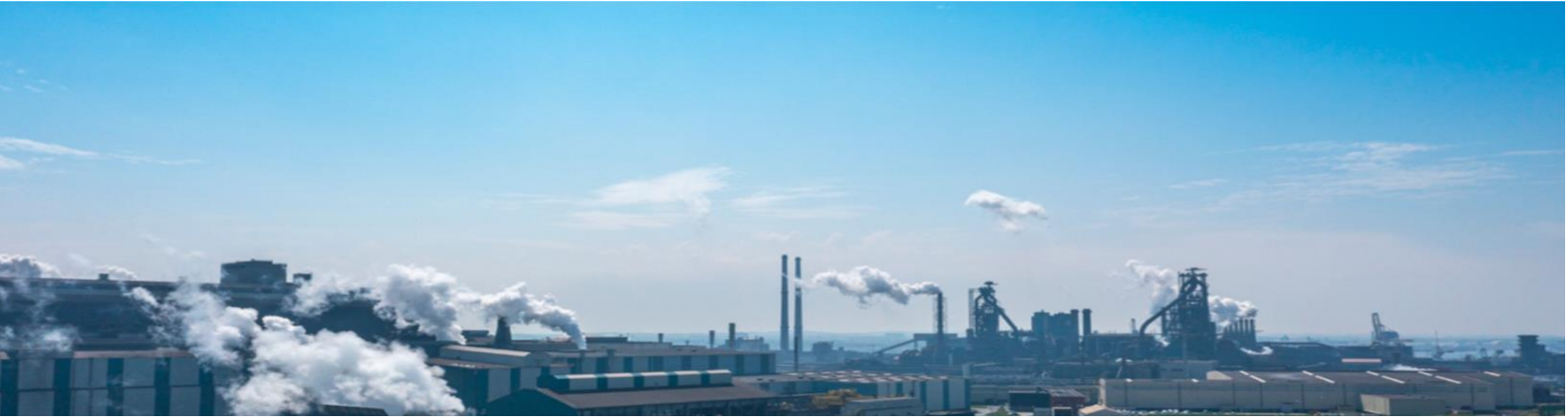

## 1. Introduction

Today's market for carbon is fragmented and complex. Limited pricing data and empirics make it challenging for investors to know existing opportunities in carbon management. The European Union (EU) constitutes one of fifteen regions worldwide that have implemented an Emissions Trading System (ETS) or a carbon tax to price carbon dioxide ($CO_2$) emissions. Due to challenges documented in the EU ETS carbon market, including international competitiveness and carbon leakage, understanding the impact of emissions pricing on these issues is critical for policy decisions in the EU, owing to the strong integration of its industries into global value chains and the region's aspirations to successfully lead climate protection. Against this background, the European Commission adopted a proposal in 2021 to establish the Carbon Border Adjustment Mechanism (CBAM). CBAM is designed to function in parallel with the EU's Emissions Trading System (EU ETS), to mirror and complement its functioning on imported goods. It will gradually replace the existing European Union mechanisms to address the risk of carbon leakage, in particular the free allowances of EU ETS. CBAM will levy a carbon price on imports of a specific selection of products to curb against 'carbon leakage' due to an ambitious climate action in Europe. China, India and the United States responded to the EU's CBAM by collectively setting a carbon price floor of US $35 per tonne of carbon emitted. The initial targeted industries for the EU's CBAM include electricity, fertilizer, cement, aluminium and iron and steel industries. Given the expected demand for managing, monitoring, pricing and reporting of carbon-transition risks that would result from the EU ETS and other global efforts under the Paris Agreement on climate change, it's clear that CBAM will increasingly play a significant role in a net-zero future.

## 2. What is Carbon Border Adjustment Mechanism?

Climate change is the defining challenge of our time (Schaeffer et al., 2025; Nyangon & Darekar, 2024; Nyangon & Akintunde, 2024; Collier et al., 2021; Hou et al., 2021). A common concern for businesses and governments when it comes to the fight against climate change is how to reduce the social cost of carbon emissions. The EU has proposed to establish the CBAM. The mechanism will attempt to curb the use of carbon by levying a steep price tag on imports of specific products (Karakaya et al., 2024). As of now CBAM will apply only to a limited number of carbon-intensive industries. These industries include cement, electricity, aluminum, fertilizer as well as iron and steel products. The free ETS allowances for EU producers will be phased out between 2026 and 2035 for the predefined industries. This phase would be transitional starting in 2023 and is expected to be fully operational by 2026.

The CBAM has been developed under the auspices of the ETS which has been in place in the EU since 2005 (Longhurst & Chilvers, 2019; Milchram et al., 2019). The EU ETS is the world's first emissions trading system that sets a cap on the amount of greenhouse gas emissions (GHG) that can be released by specific industries under the ETS allowances or bought and exchanged on an open market. However, a major challenge that the EU ETS has grappled with over the years is how to provide the number of ETS free allowances in order to mitigate potential carbon leakage. Carbon leakage refers to the increase in emissions when an industry relocates from a country with stringent climate rules to a country with weak environmental regulations. The CBAM also falls under the umbrella of the EC Green Deal whole of society approach to fighting climate change. As part of the Green Deal initiative, the EC committed to achieve climate neutrality by 2050. The CBAM is a response to a realization that a lack of ambition of third countries could undermine EU climate mitigation efforts. This would expose the EU to a risk of direct carbon leakage and declining EU competitiveness in energy-intensive industries. Direct carbon leakage would also occur in the EU when companies transfer their production to countries that are less strict about limiting emissions. The CBAM seeks to counteract this potential risk by providing more structure to the ETS market, i.e. putting a carbon price on the imports of certain carbon-intensive goods sourced from outside the EU and potentially rebate the carbon price already paid for through the EU exports (Nyangon, 2017).

Carbon border adjustment programs are becoming more predominant. In the United States, there are two ongoing programs in California and through the Regional Greenhouse Gas Initiative (RGGI). California's cap and trade system covers approximately 85% of the region's GHG emissions. The California's carbon border adjustment system is applied exclusively to its power market. On the other hand, RGGI covers 11 northeastern states of Connecticut, Delaware, Maine, Maryland, Massachusetts, New Hampshire, New Jersey, New York, Rhode Island, Vermont and Virginia, and has been in place since 2009 (Nyangon & Byrne, 2023, 2018; Rissman et al., 2020). The RGGI sets a general regional cap on emissions originating from power generation at the federal level. Pennsylvania has applied for RGGI membership and North Carolina is considering joining the initiative.

## 3. Stylized Facts of CBAM and EU Emissions by Country and Industry

### 3.1. Empirics, Analytics and Opportunities

Is there a role for analytics in improving optimization and forecasting of ETS processes, reduce some of the market inefficiencies that could jeopardize objectives of CBAM? For instance, to meet the 2030 target, European steelmakers have significant low-carbon projects in the pipeline but haphazardly implemented and weakly-monitored CBAM could drive up the industries regulatory costs thereby jeopardizing the industry's investment in green steel as well as weaken existing carbon leakage protection (Zare et al., 2024). The following opportunities for analytics exist in mitigating the risk of carbon leakage and their implications for the ETS value chain, including:

a) Carbon credits validation and verification using artificial intelligence (AI) and machine learning (ML) embedded tools using satellite image, historic benchmarks, etc. (Nyangon, 2025a).
b) Planning and forecasting analytics tools for ETS developers to evaluate new projects based on project type, area, region, co-benefits etc.
c) Carbon accounting and reporting solutions includes analytics infrastructure to track and report emissions from the five carbon intensive industries (i.e., electricity, fertilizer, cement, aluminium smelters, and iron and steel) in addition to analytics and consulting services around reducing said emissions.
d) Modernization of CBAM infrastructure with growing use of connected devices across supply chains, improving data collection and transmission.

e) Firms looking to achieve net zero targets via offset purchases seek an understanding of the market and its dynamics. Use of analytics to better understand the CBAM market and its dynamics, including modeling and forecasting future ETS market to help balance the stream of the ETS allowances issued.

f) ETS allowances are priced based on a myriad of factors, including energy types and prices (oil, coal, electricity, natural gas), economics (industrial production, economic sentiment), extreme weather, unanticipated temperature changes (precipitation), and certified emission reduction (CER), and marginal abatement cost (MAC). The ability to model these factors and variety of data sources using advanced analytics technologies would be powerful in streamlining the pricing process (Batel & Rudolph, 2021; Laes et al., 2014; Lee et al., 2020).

g) Building a central data and analytics platform to understand the ETS market trends, supply chain mechanics to recognize high-emitting processes and embedded carbon in products, and facilitate transparency would increase liquidity and CBAM market functionality.

## 3.2. Carbon Leakage

The CBAM could be seen as a replacement for the free ETS allowances currently granted to EU producers believed to be at high risk of carbon leakage. It will be applied to imports at the price of carbon set by the EU ETS system through auctions, which are expected to increase over time as free allowances - initially only in the five industries that were described above but potentially later in other import competitive industries - that are already subject to the EU ETS (including ceramics, glass, paper and other chemicals) (Nyangon et al., 2017; Siddiqui et al., 2019; Sakai & Barrett, 2016). As we are writing this paper, the EU carbon price is currently already approaching the threshold of €80 per tonne of $CO_2$, is it perfectly possible that the effective price of carbon paid by EU producers in industries which are subject to a lot of import competitiveness covered by the EU ETS, will gradually increase to a higher level by the end of 2035, which is likely to lead to carbon leakage in these countries and therefore justifies the introduction of the CBAM. Figures 1 and 2 illustrate carbon dioxide emissions per industry for the EU 27.

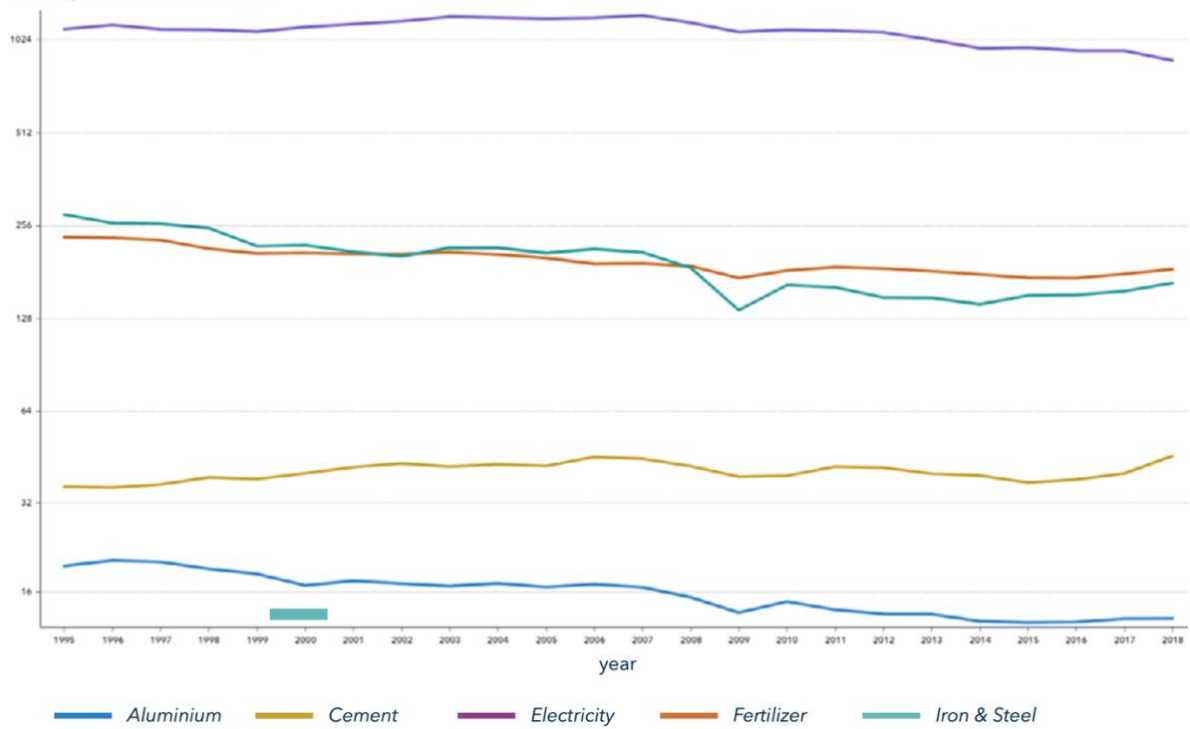

*Figure 1: Logarithmic estimation of annual carbon dioxide emissions per industry (tons of $CO_2$)*

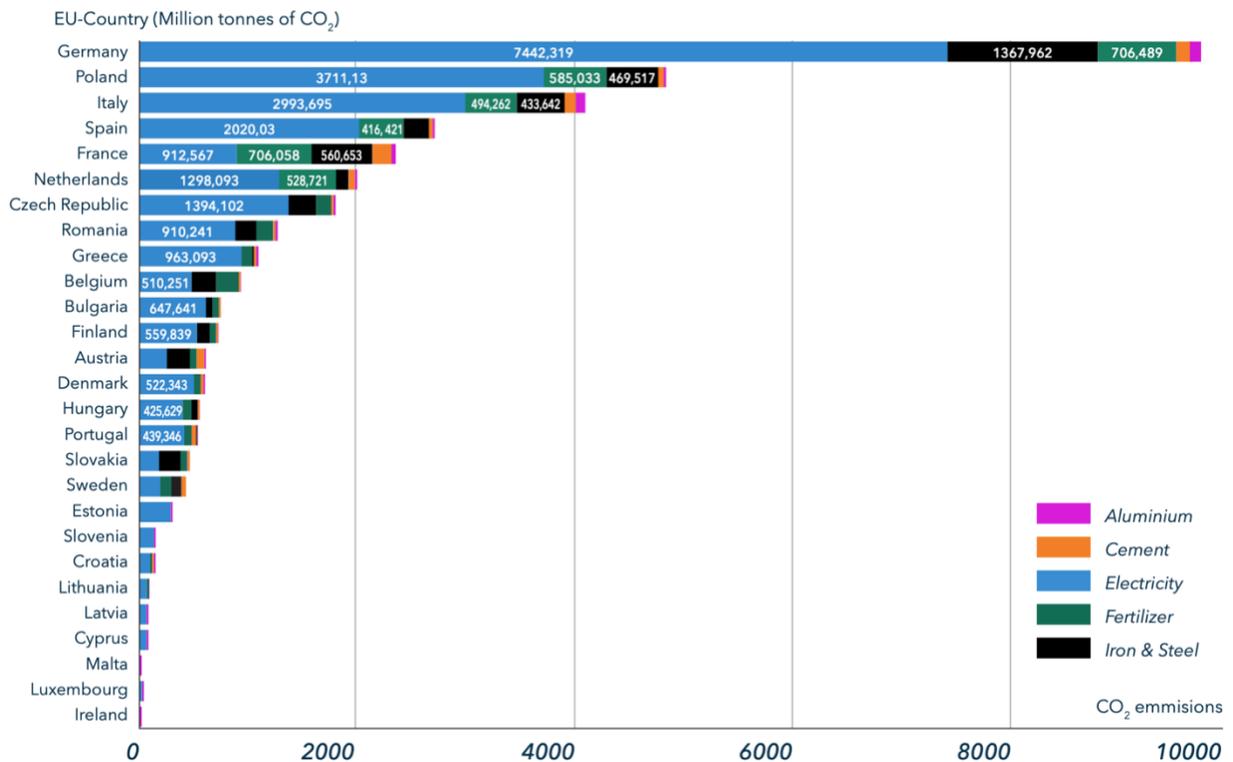

*Figure 2: Carbon dioxide emissions per industry for EU 27 countries (tons of $CO_2$)*

Having determined that the phasing out of free ETS allowances will significantly increase the price of carbon faced by importing competitive carbon-intensive producers in the EU – the commission proposes that the parallel implementation of the CBAM is therefore a measure that could be used in order to avoid carbon leakage.

The original proposal applies to imports of the five industries mentioned above. EU importers of non-EU products will have to pay a levy. EFTA countries (Iceland, Liechtenstein, Norway and Switzerland) are exempt because of their participation in or linkage to the EU ETS. Other countries with comparable CO2 prices could follow in the future. The regulation focuses on direct emissions from the production process (scope 1), although the scope could be extended to purchased electricity (scope 2) and other upstream emissions (scope 3) after the transition period. Important to stretch here is that the price of the CBAM certificates will reflect EU ETS prices, adjusted for any free allowances EU producers still receive and carbon costs incurred during the production process in the producing country.

The CBAM also contains an important article on circumvention. The EU will monitor significant changes in trade flows and slightly modified products that "have insufficient justification or economic justification other than to avoid obligations under this Regulation". Thus, the Commission's proposal empowers the Commission to react in cases of 'circumvention'. These are situations where a change in a trade flow pattern has insufficient economic justification other than avoiding the obligations of the CBAM. For example, if the pattern of trade to the EU from a third country would show a marked decrease in the goods covered by Annex I (such as iron pipes) and a related increase of downstream goods that use the Annex I goods covered by the CBAM as inputs (such as tables using such iron pipes for framing), the Commission may act accordingly. Such circumvention related action could extend the CBAM's scope.

## 4. Carbon Intensive Industries

### 4.1. Cement

When we look at the European environmental performance of cement in manufacturing, the natural execution of cement fabricating is generally homogenous, given that almost 60% of outflows stem from the calcination process that converts limestone to quicklime. Contrasts inside Europe basically emerge from the fuels utilized to create heat in cement ovens. There are alternative heating technologies however that use power or hydrogen that can help to decarbonize. Most European cement companies have already headed in the right direction by going from wet to less energy-intensive dry production methods. Carbon capture and sequestration will subsequently be a fundamental component in any strategy towards full decarbonization of the cement industry.

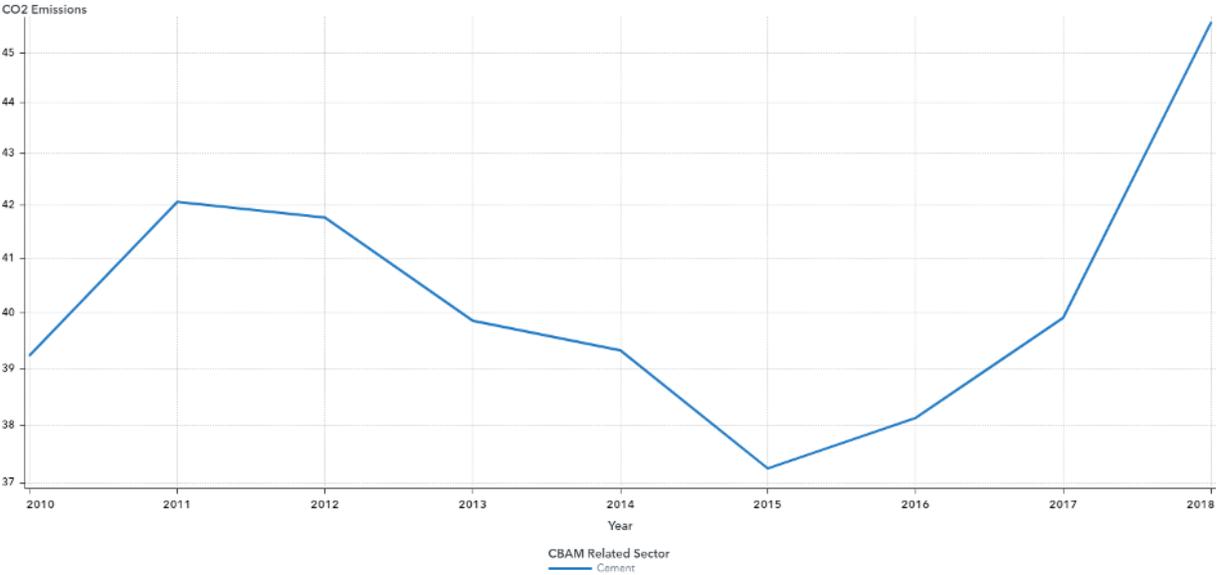

*Figure 3: Cement industry*

### 4.2. Electricity and Power

Electricity as such can be considered as a homogenous asset due to physical characteristics. Nonetheless we must consider the reliability of the supply which defines the different market segments (Nyangon, 2025b). Another important key differentiator is origin and more specifically the clear distinction between renewable and non-renewable energy sources - accompanied by a significant difference in their respective carbon intensity.

The power industry on the other hand is heterogenous. State owed enterprises used to dominate here, but it has been decentralized into many parts like distribution, transmission, generation where a system operator is guaranteeing a supply/demand balance.

According to Eurelectric, the sector association representing common interests of the electricity industry at a European level, achieving a cost-effective pathways for a climate neutral European economy with net-zero greenhouse gas emissions by 2050 will depend highly on four key factors: a) electricity supply with over 80% from renewables, b) diversification of power sources to ensure system reliability and flexibility, c) changing role of conventional generation, which will provide back-up energy while gradually being less used for energy production; d) maturity

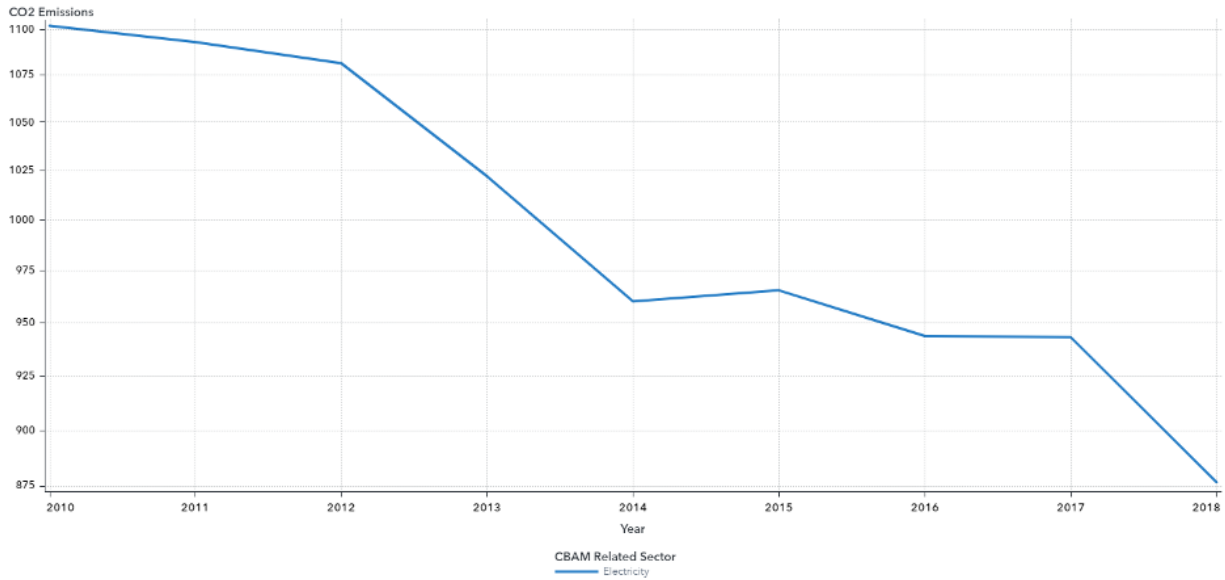

of $CO_2$ offset and power-to-gas technologies (Matijašević et al., 2022; Ruiz et al., 2023).

*Figure 4: Electricity industry*

### 4.3. Iron and Steel

There are in essence two processes which lead to the generation of crude steel. The primary production starts with iron ores where hot metal is produced and as a next phase is being converted to crude steel in a basic oxygen oven. The secondary production merely is produced by the smelting of scrap (or instantaneously reduced iron in an electric arc oven. The two processes are fundamentally different in the following aspects: necessary energy, metallurgical process, output of the produced steel (both quality and destination of use) and the process emissions.

When we compare the Carbon intensity of steel production, we see clearly that the amount really depends upon the production process. Primary steel production is the most carbon intense process - as it needs the most energy input. Secondary production on the other hand is relatively less carbon intensive when transforming scrap steel. Why do the $CO_2$ emissions exceed so much? Most importantly you have the fuel combustion in the coking and sintering process, hot metal production and finally the conversion to steel. To give some idea about the global average carbon release: for every ton of steel that has been produced, 2.3 tons of $CO_2$ are released during the steelmaking process. European steelmakers have on average 17% lower CO2 (**1.9** tons) per ton of steel.

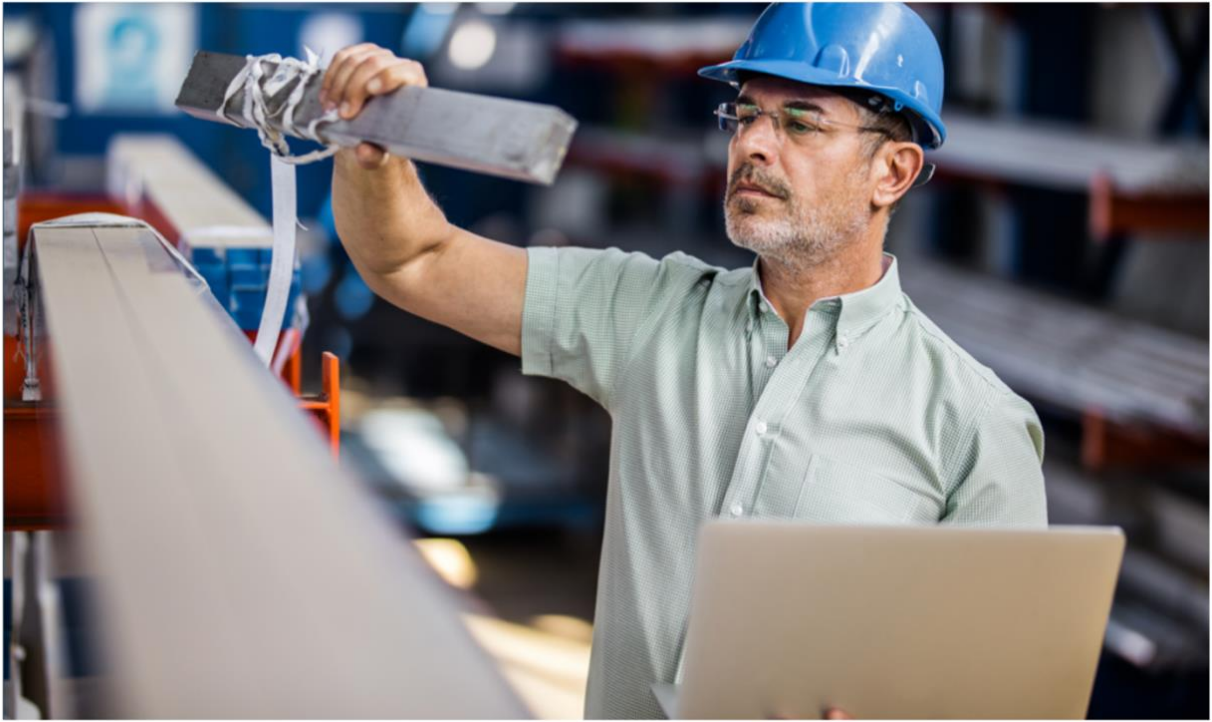

## 4.4. Fertilizers

As an industry within the chemicals sector, fertilizers are considered as a separate entity due to their economic importance. Next to this, the fertilizer industry is defined by a couple of characteristics that differentiate it from the chemical sector as a whole. Fertilizers are typically made up of three main elements: nitrogen, phosphate and potassium. Of those three elements, nitrogen contains the most significant GHG emissions, and it is the only element of the fertilizer industrial complex to be covered under the EU ETS. Two activities under the broad heading of fertilizers have ETS benchmarks: production of ammonia and production of nitric acid. A feedstock of natural gas is used in the production of ammonia, from which hydrogen is broken out with steam and pressure. At the same time, the most $CO_2$ intensity is happening during this phase of the process. The reason being that energy is mainly consumed here and because of the stream of $CO_2$ produced in the process of hydrogen conversion to ammonia.

Most of the time natural gas is being used as a feedstock to hydrogen production at EU level. As gas is not as abundant in Europe compared to countries like Russia - EU producers pay relatively more. Due to this fact, European producers have been incentivized to be more efficient and reduce the amount gas. The $CO_2$ emissions per ton of ammonia produced in the EU range is on average 1.9. While countries like Russia and China are CO2 intensive with a respective average of 2.4 and 5 tons.

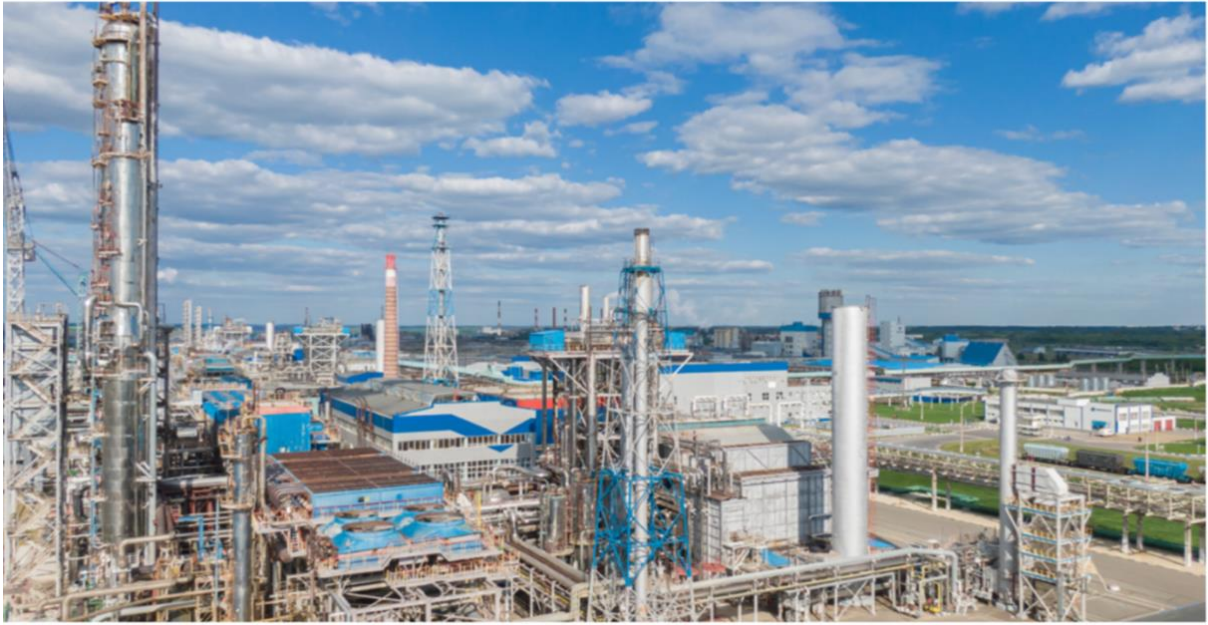

### 4.5. Aluminium

Aluminium is a metal that is relatively abundant. In the primary production process, aluminium is produced from the ore bauxite, which is purified to yield aluminium oxide –also known as alumina – and reduced to elemental aluminium in smelting plants through an electrochemical process, which requires temperatures in excess of 950°C and a high intensity electrical current. Secondary aluminium is refined or remelted from scrap metal recovered from waste and recycling streams, requiring a melting furnace operating at temperatures ranging from 700°C to 760°C, mostly using natural gas.

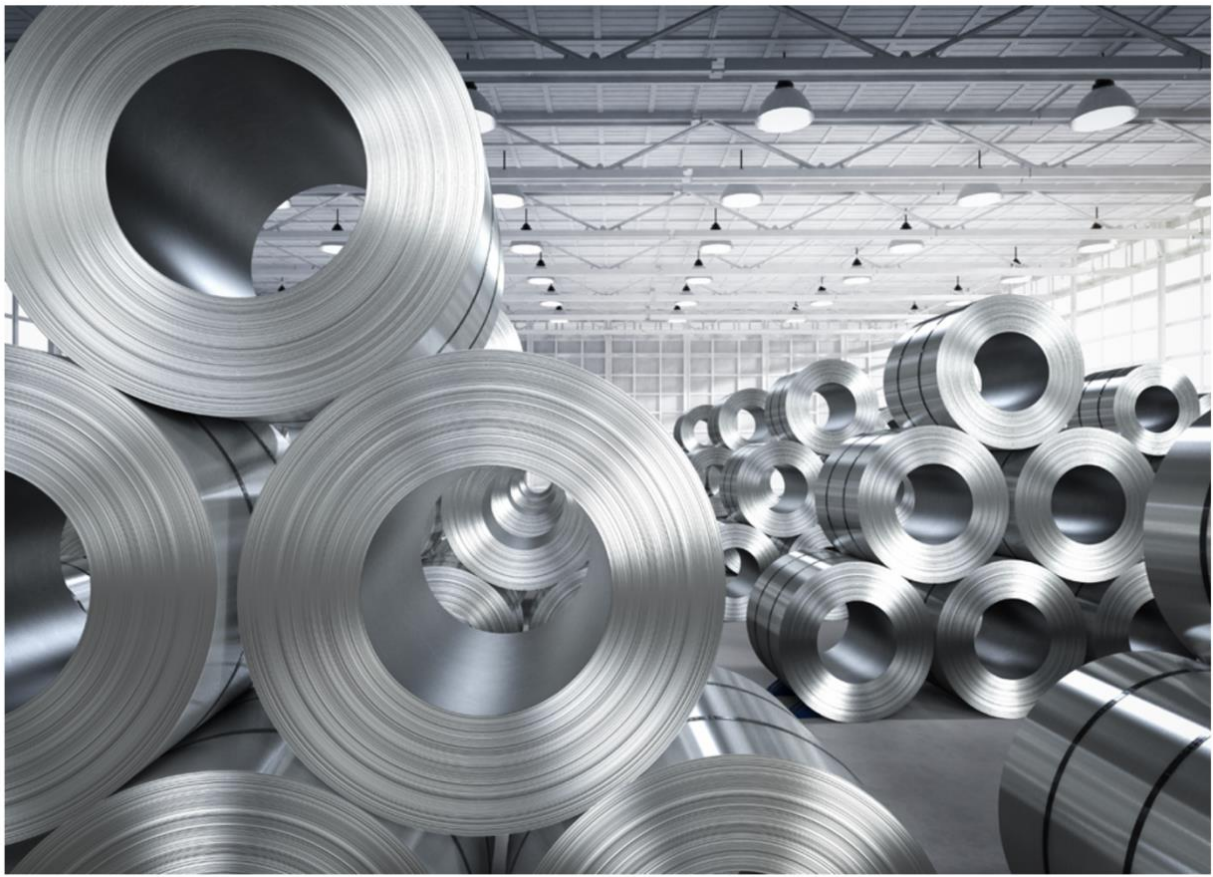

## 5. Econometric Analysis of Carbon Leakage

Do countries with relatively weak environmental regulation attract pollution intensive production? The carbon leakage problem has been an econometric research question for more than 40 years. Recent studies employing advanced methodologies show statistically significant evidence of pollution haven. Most of these studies recommend three main approaches for mitigating carbon leakage, namely (i) compensating for excessive costs through free allocation of emission allowances, (ii) leveling the playing field of border carbon adjustments, and (iii) directly fostering low-carbon innovation. Any unpaid leakage instrument that a jurisdiction may wish to use needs to be compatible with broader relevant trade regimes such as the World Trade Organization (WTO) and EU trade rules. This is an important consideration weather the ETS is linked to another ETS or not. When the market is linked in one market using partial or full convergence of allowance prices, the ETS linking removes or reduces any internal competitive distortion caused by differences in the two markets.

The empirical literature on carbon leakage is undergoing change as both carbon border adjustments and innovation policies gain relevance as anti-leakage approaches. For example, the carbon anti-leakage approaches relying on linear (logarithmic) approximations and predictions of carbon stringency policies could be problematic. Because the incentives for production relocations increase over proportionately with the stringency of the policy measure in the presence of fixed costs of relocation, dynamic approximation approaches are needed.

To estimate the effects of the CBAM on individual products, we adopted the following rate calculation:

$$\text{CBAM}_{i,j} = (\text{CP}_{EU} - \text{CP}_i) \times \beta_{i,j} \tag{2}$$

Where:

$\text{CBAM}_{i,j}$ – represents CBAM rate for imports of $j$ industry originating form country $i$

$\text{CP}_{EU}, \text{CP}_i$ – represents carbon price in the EU and in country $i$

$\beta_{i,j}$ – represents emission intensity in $i$ and $j$ industry

$i$ – represents countries while $j$ represents the five CBAM priority industries

The design of an effective, legitimate and fair EU CBAM is fundamental to tackling the issue of carbon leakage more efficiently, ensuring full compatibility with the WTO rules and compliance with the Paris Agreement goals .The econometric modeling approach above could take four major forms, namely (i) a carbon tax on selected high emission products exposed to carbon leakage effects; (ii) a tariff imposed on imports at the EU border on carbon-intensive products; (iii) obligatory purchase of emission permits by external importers from a designated EU seller at rates similar to those offered by the EU ETS; and (iv) extension of the existing EU ETS taxation in the form of emission permits imposed on foreign imports.

## 6. The EU CBAM and Analytics Opportunities

The current proposal suggests that, once the CBAM is in force and the Commission adopts the necessary implementing legislation, it may also have an indirect coverage extending to many other products and sectors than those formally covered. As stated before, the CBAM proposal targets those sectors where the risk of carbon leakage is high – while at the same time falling under the scope of the ETS, and where there is the possibility to calculate embedded GHG emissions. An interesting observation is that in the current CBAM proposal, the Commission also leaves the room for indirect coverages extending to many other products and sectors than those formally covered. In the Annex I of the proposal - the European CN (Combined Nomenclature) codes of the exact types of goods that would be covered are listed for clarification.

The next 4 four types of complex goods would also fall under CBAM:

- white Portland cement [CN 2523 21 00]
- Mineral or chemical fertilisers containing the three fertilising elements nitrogen,
- phosphorus, and potassium [CN 3105 20]
- Tubes and pipes, having circular cross-sections, the external diameter of which
- exceeds 40.64 cm, of iron or steel [CN 7305],
- Aluminium foil [CN 7607]

But what about the embedded emissions of goods falling outside the industrial sectors explicitly covered (iron and steel, aluminium, cement, fertilizers) but inside the CN Codes listed in the proposal's Annex? Manufacturing the complex goods listed, such as the examples above, will require a number of input materials, some of which may be outside the scope of Annex I. Let's take the manufacturing process of aluminium foil; this requires other elements than aluminium alone like biocides, paper or other chemicals.

What still needs to be clarified in the EC's proposal is the definition of the boundaries for the calculation of embedded emissions for complex goods from products in other categories (e.g., paper and biocides) that are included in calculating the emissions from a complex good falling within the CN Codes of Annex I (e.g., aluminium foil).

In order to bring clarity, the EC will need to set "system boundaries" that define the exact input materials whose embedded emissions are to be added when calculating the total emissions of a complex good covered by the CBAM. By doing so, the EC can precisely determine the categories of goods that the CBAM indirectly covers as well.

One can see that this first transition phase already with a limited CBAM scope can quickly become quite complex. A good data strategy will be essential – presuming that the scope of CBAM will be subject to change every year (e.g. new sectors that will be covered, CO2 conversion rates, etc.).

Countries with no national carbon pricing scheme such as the US, Russia, Turkey, China, and India are most likely to be significantly affected by the EU CBAM. The EC's proposal empowers it to take action in cases of circumvention of the CBAM. These are cases where there is a change in the pattern of trade which has insufficient economic justification other than avoiding the obligations of the CBAM. Let's take the example of iron pipes, if the trade pattern from a third country to the EU would show a significant decrease in the iron pipes (or other goods covered in Annex I) while at the same time a related increase of downstream goods like tables using such iron pipes for framing, the EC may act. These potential circumventions often cannot be spotted by the naked eye. Trade history teaches us that some companies are quite creative by applying subtle changes to avoid sanctions. This will be a new challenge for policy officers who seek to protect the common EU market and are confronted with a new data source that will scale fast.

# 7. Conclusion

The EU CBAM is a policy measure in the form of tariffs or allowances that proposes to price carbon at the border for certain imports. This would change the dynamics of international trade in carbon-intense products, including fossil fuels. The EC's proposal defines the scope of the CBAM to ensure its integrity and effective alignment with the EU ETS and the EC's "Fit for 55" plan for a green transition - i.e. EC's measures to reduce GHG emissions by 55% in 2030 from their 1990 levels. The EU CBAM proposal is designed to apply to specific industrial sectors for which the EU domestic industry is subject to the ETS, at high risk of carbon leakage, and for which it is feasible to calculate embedded GHG emissions. The EC will need to set "system boundaries" that would define the exact input materials whose embedded emissions must be added when calculating the total emissions of a complex good covered by the CBAM. These boundaries will also be important for the circumvention of sanctions and the shifts in trade to downstream goods in order to avoid the CBAM. To uncover and understand these insights, EC needs to invest in the next generation of advanced analytics and AI for better decisioning to protect domestic companies against polluting firms from outside of the EU region to help tackle climate change. For effective implementation of the CBAM and other similar carbon mechanisms, the EU needs to forge international alliances around carbon pricing with other regional trading blocks - ASEAN, APEC, BRICS, USMCA, CIS, COMESA, SAARC, MERCOSUR, IOR-ARC, and others. When practical, CBAM could level the playing field between domestic producers and foreign suppliers in high-emission industries from these trading blocks.

17